\documentstyle[aps,prl]{revtex}
\begin{document}
\input epsf.sty
\twocolumn[\hsize\textwidth\columnwidth\hsize\csname %
@twocolumnfalse\endcsname
\draft
\widetext
%%%%%%%% prl (above) %%%%%%%%%%%%%%%%%%

\title{Nanoscale anisotropic structural correlations in the paramagnetic
and ferromagnetic phases of Nd$_{0.5}$Sr$_{0.5}$MnO$_3$}

\author{V. Kiryukhin$^1$, B. G. Kim$^1$, T. Katsufuji$^2$,
 J. P. Hill$^3$, and S-W. Cheong$^{1,2}$}
\address{(1) Department of Physics and Astronomy, Rutgers University,
Piscataway, New Jersey 08854}
\address{(2) Bell Laboratories, Lucent Technologies, Murray Hill,
New Jersey 07974}
\address{(3) Department of Physics, Brookhaven National Laboratory, Upton,
New York 11973}

\date{\today}
\maketitle

\begin{abstract}

We report x-ray scattering studies of short-range structural 
correlations and diffuse scattering in Nd$_{0.5}$Sr$_{0.5}$MnO$_3$. 
On cooling, this material undergoes a series of transitions, first from
a paramagnetic insulating (PI) to a ferromagnetic metallic (FM) phase, and 
then to a charge-ordered (CO) insulating state.
Highly anisotropic structural correlations
were found in both the PI
and FM states of this compound. 
The correlations increase with decreasing temperature, reaching a maximum
at the CO transition
temperature. Below this temperature, they
abruptly collapsed.
Single-polaron diffuse scattering was also observed in both the PI  
and FM states suggesting that substantial local lattice distortions
are present in these phases.
We argue that our measurements
indicate that nanoscale regions exhibiting layered 
orbital order
exist in the paramagnetic and ferromagnetic phases of
Nd$_{0.5}$Sr$_{0.5}$MnO$_3$.

\end{abstract}


%%%%%% prl format (below) %%%%%%%%%%%%%%
\phantom{.}
]
\narrowtext
%%%%%%%%% prl (above) %%%%%%%%%%%%%%%%%%

Manganite perovskites of the chemical formula A$_{1-x}$B$_x$MnO$_3$ (where
A is a rare earth, and B is an alkali earth atom) have recently attracted 
considerable attention because they exhibit a number of interesting electronic
properties, including the colossal magnetoresistance 
phenomenon \cite{Review}. 
Recent studies show convincingly that a number of the intriguing properties of
the manganites cannot be understood through spatially uniform 
phases and that local inhomogeneities must play an essential role in these
compounds \cite{Local}.
The high resistivity of the paramagnetic insulating (PI) phase
found in most of the manganites at high temperatures, for example, results in 
part from the presence of the lattice polarons that form when an $e_g$ electron
localizes on a Mn$^{3+}$ site, inducing the Jahn-Teller
distortion of the MnO$_6$ octahedron \cite{Millis}.
Such lattice polarons were recently
detected directly in the PI phase of the manganites by means of diffuse
x-ray and neutron scattering \cite{Polarons,Corr}. 
Magnetic and lattice polarons have also been found in other
phases of the manganites, including the ferromagnetic metallic (FM)
and insulating, 
and antiferromagnetic insulating phases \cite{Corr,Hen}.

The properties of the high-temperature paramagnetic insulating phase were
recently found to exhibit sharp anomalies at several commensurate concentrations
of doped carriers, such as $x$=3/8 and $x$=1/2 \cite{Kim}. 
The presence of these anomalies
suggested that local
charge or orbital correlations of some kind, possibly 
short-range polaron-polaron correlations, play an important role in the PI
phase. Recently, such polaron correlations were indeed observed in 
La$_{1-x}$Ca$_x$MnO$_3$, La$_{2-2x}$Sr$_{1+2x}$Mn$_2$O$_7$, and 
Pr$_{1-x}$Ca$_x$MnO$_3$ manganites \cite{Polarons,Corr}. 
Interestingly, the high-temperature
structural correlations do not necessarily correspond to the low-temperature
order,  and in some cases the incipient high-temperature ordering 
is in direct competition with the ground state \cite{Corr}.
The relationship between the low-temperature
ground state and the high-temperature correlations
is at present
largely unclear. 

While local structural correlations appear to play an important role in the 
macroscopically homogeneous states of the manganites, the
microscopic nature of these correlations, 
as well as the extent to which they affect the
electronic properties of the manganites, have yet to be established.
To answer these questions, extensive experimental and theoretical work
is needed.

In this paper, we report  
synchrotron x-ray scattering studies of short-range structural correlations
and diffuse scattering in Nd$_{0.5}$Sr$_{0.5}$MnO$_3$. We find a new type
of correlated anisotropic lattice distortion
in the ferromagnetic and paramagnetic phases
of this material. We argue that these correlated distortions
are associated with local regions of layered orbital and, possibly,
magnetic order.
These observed structural
correlations then provide a natural explanation for the anisotropic
magnetic properties of this three-dimensional perovskite compound.

Single crystals of Nd$_{0.5}$Sr$_{0.5}$MnO$_3$ were grown using the standard
floating zone technique. The x-ray diffraction measurements were carried out
at beamlines X22A and X20C at the National Synchrotron Light Source. 
In each case,
the x-ray beam was focused by a mirror, monochromatized by a Si (111)
monochromator, scattered from the sample mounted inside a closed-cycle
cryostat, and analyzed with a pyrolytic graphite crystal. In this paper,
Bragg peaks are indexed in the orthorhombic {\it Ibmm}
notation in which the longest
lattice constant is $c$.

Nd$_{0.5}$Sr$_{0.5}$MnO$_3$ is a paramagnetic insulator at high temperatures.
With decreasing temperature, it undergoes a transition to a ferromagnetic 
metallic state at T$_c$=250 K,
and at T$_{CO}$=150 K, it becomes
a CE-type charge ordered insulator possessing the 
$d_{3x^2-r^2/3y^2-r^2}$
type orbital ordering \cite{NdSr}. In addition, 
weak magnetic Bragg peaks due to
the presence of an
A-type antiferromagnetic impurity phase, in which ferromagnetic layers are
stacked antiferromagnetically along the (001) direction,
have been observed in 
some samples below T$_{AF}$=200 K \cite{AF}. 
Finally, in some of the samples studied
previously, these phases were found to coexist with the higher-temperature
phases either below T$_{AF}$ or below T$_{CO}$ \cite{Vogt}. 
Note, these samples were all found
to consist of a single structural phase at temperatures higher than 200 K. 

\begin{figure}
\centerline{\epsfxsize=2.9in\epsfbox{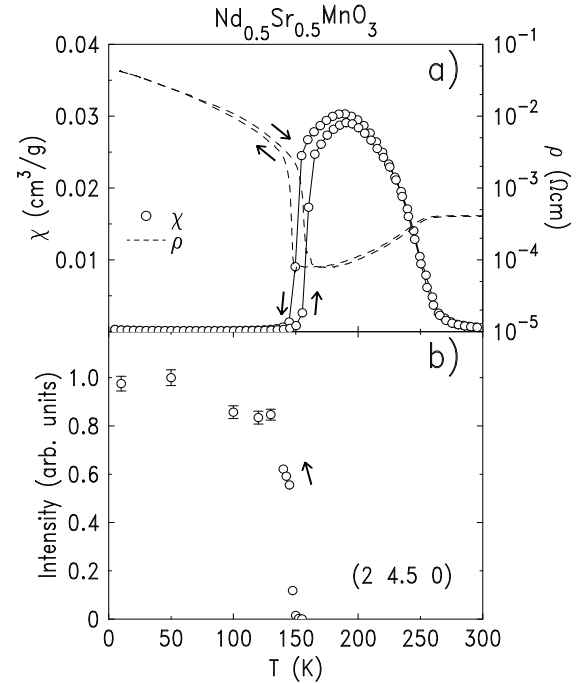}}
\vskip 5mm
\caption{(a) Temperature dependence of the zero-field electrical resistivity 
(dashed line) and
the magnetic susceptibility (open circles) in a
magnetic field of 2000 Oe. (b) The
intensity of the (2, 4.5, 0) peak characteristic of the CE-type charge and
orbitally ordered state.}
\label{fig1}
\end{figure}

The bulk properties of our sample are consistent with the results 
published in the literature. Fig. \ref{fig1} shows the temperature dependence of
the magnetic susceptibility, electrical resistivity,
and the intensity of the (2, 4.5, 0) peak, which is
characteristic of the CE-type charge and orbitally ordered state.  
From the data of Fig. \ref{fig1}, we obtain T$_c\approx$260 K, and
T$_{CO}\approx$150 K. There is a measurable hysteresis in the magnetic
susceptibility below T$\approx$200 K, and therefore it is possible that the
sample contains several structural phases at low temperatures, as was 
reported in Ref. \cite{Vogt}.

\begin{figure}
\centerline{\epsfxsize=2.9in\epsfbox{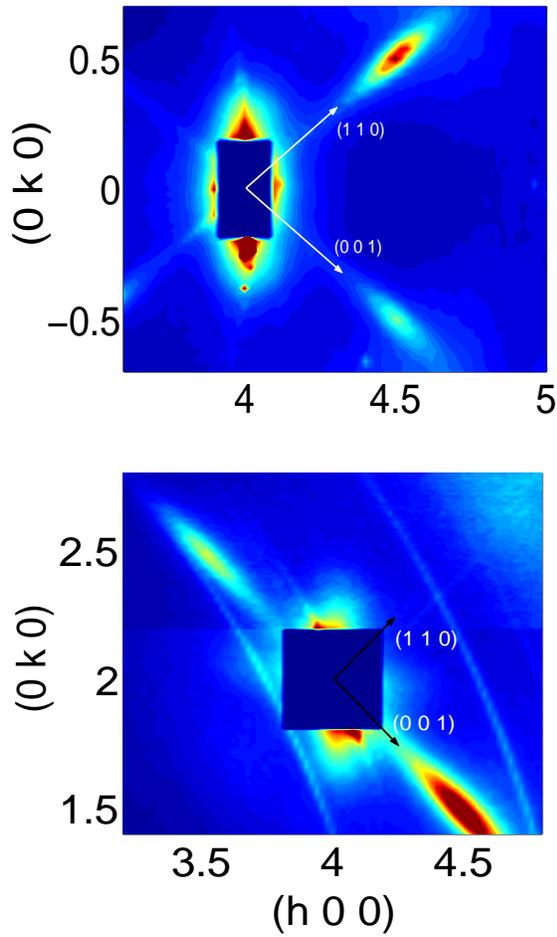}}
\vskip 5mm
\caption{(Color) Contour plots of the scattered
x-ray intensity around the (4, 0, 0)
Bragg peak at T=165 K (top panel), and around the (4, 2, 0) peak at T=270 K
(bottom panel). (See the text for the explanation of the effects of twinning.)
The two light circular arcs in the bottom panel are due to impurity powder
scattering.} 
\label{fig2}
\end{figure}

In order to characterize the short-range structural correlations and the 
single-polaron diffuse scattering, we have measured the x-ray intensity in the
vicinity of several representative Bragg peaks. Fig. \ref{fig2} shows
a contour plot of the x-ray intensity in the vicinity of 
the (4, 0, 0) and the (4, 2, 0) Bragg points. The crystals are twinned in
all the three cubic directions of the underlying cubic perovskite structure,
and therefore the data of Fig. \ref{fig2} consist of a superposition
of the scattering due to all the twin domains in the crystal. In particular,
in the top panel of Fig. \ref{fig2}, the
Bragg reflections (4, 0, 0) and (2, 2, 4) coincide,
and scattering from both the (hk0) and the (hhl) zones is present.
Similarly, the Bragg reflections (4, 2, 0), (3, 3, 2),
and (1, 1, 6), and combined scattering from the (hk0) and (hhl) zones are
present in the bottom panel of Fig. \ref{fig2}. The main coordinate axes in this
figure show the (hk0) zone; the 
axes for the (hhl) zone are superimposed on the figure. 

\begin{figure}
\centerline{\epsfxsize=2.9in\epsfbox{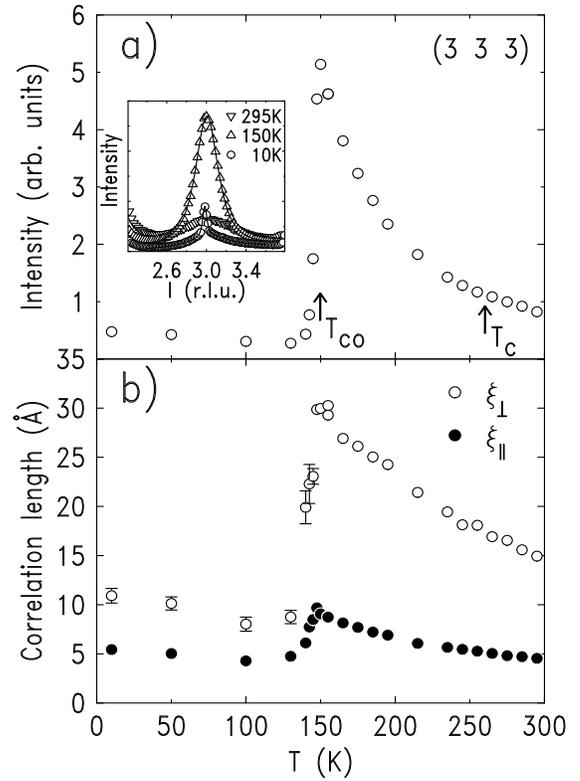}}
\vskip 5mm
\caption{Temperature dependence of the x-ray intensity of the (3, 3, 3)
peak (a), and of the correlation lengths of the corresponding structural
fluctuations (b). $\xi_\perp$ is the correlation length perpendicular to the
reduced scattering vector (001) and $\xi_\parallel$ is the correlation length
parallel to this vector (the ``in-plane'' and ``out-of-plane'' correlation
lengths). The inset in (a)
shows scans through the (3, 3, 3) peak, along the (00l) 
direction at different temperatures.} 
\label{fig3}
\end{figure}

The data of Fig. \ref{fig2} exhibit intense,
anisotropic, and very broad scattering at the
(4.5, 1.5, 0), (3.5, 2.5, 0), and (4.5, $\pm$0.5, 0) points
in the (hk0) axes setting \cite{Abs}. 
Alternatively, this scattering can be ascribed to the (3, 3, 3), (3, 3, 1), and
(2, 2, 5) reciprocal space coordinates. For a reason that will become
clear later, we will
use the latter assignment, in which the scattering occurs with a (0, 0, 1)
reduced scattering vector. 
Note, for crystal with {\it Ibmm} symmetry, as Nd$_{0.5}$Sr$_{0.5}$MnO$_3$
has at temperatures higher than T$_{CO}$
\cite{NdSr,Vogt}, the (3,3,3), (3,3,1), and (2,2,5)
reflections are forbidden.
We have studied the temperature dependence of the
scattering at these points. Scans were taken along the (00l), (hh0), and (h-h0)
directions. The data collected in the (hh0) and (h-h0) scans were
essentially identical, and in what follows we will not distinguish them,
simply referring to them as
scans perpendicular to the (00l) direction.
Examples of the scans in the (00l) direction
at the (3, 3, 3) reciprocal lattice point are shown in the inset in
Fig. \ref{fig3}. To parametrize the data, the scans were fitted to
a Lorentzian to the power 1.5 line-shape convoluted with the instrumental
resolution, and the resulting fits were used to extract the peak intensities and
the intrinsic peak widths.

Because the diffracted x-ray intensity is proportional to the square of the 
Fourier transform of the atomic density-density correlation function,
the broad peaks of Fig. \ref{fig2} signal the presence of correlated lattice
distortions above T$_{CO}$. The anisotropic correlation
length (the domain size) of these regions can be extracted from the
intrinsic width of the peaks and is shown in
Fig. \ref{fig3} along with
the temperature dependence of the x-ray intensity at the
(3, 3, 3) reciprocal lattice point.
We find that the correlation length is
approximately 3 times smaller in the (001) direction than in the perpendicular
direction at all temperatures. 
As the temperature decreases, both correlation lengths grow and
reach their maximum at a temperature just above T$_{CO}$. The scattering then 
abruptly collapses at T$_{CO}\approx$150 K. 
Below T$_{CO}$, two components in the
scattering are
observed, a sharp component
and a broad component (inset in Fig. \ref{fig3}). The 
integrated intensity of the sharp component was approximately 100 times smaller
than the integrated intensity of the correlations
measured at T=150 K, and was approximately
independent of temperature below T=150 K.
The possible
multiphase character of the low-temperature state makes it hard to assign
these observed components to specific phases.

The large anisotropy of the structural correlations observed above 
T=150 K makes it
natural to propose that they correspond to the lattice distortions arising from
the presence of some kind of layered orbital order in the correlated domains.
One possible scenario is that these correlated domains possess the layered
$d_{x^2-y^2}$-type orbital order \cite{NdSr} and exhibit the associated A-type
antiferromagnetism. This scenario is supported by a recent observation
of anisotropic spin fluctuations in Nd$_{0.5}$Sr$_{0.5}$MnO$_3$ by means
of neutron scattering \cite{AF}. These spin fluctuations exhibited 
two-dimensional character in the PM phase, and were of the antiferromagnetic
A-type in the FM phase \cite{AF}.
In addition, Nd$_{1-x}$Sr$_x$MnO$_3$ samples with $x>$0.51 exhibit the A-type
antiferromagnetic metallic (AFM) low-temperature state, and therefore strong
AFM fluctuations can be expected in the $x$=0.5 compound which lies close to
the AFM phase boundary.
In the above picture of the correlations,
the reduced wavevector of the structural distortion characteristic
of the correlated domains in our sample, (0, 0, 1), is
perpendicular to the FM layers, and the periodicity of the lattice modulation
is twice the distance between these layers \cite{Explanation}.
 
Is is also possible that the structural correlations of Fig. \ref{fig2}
reflect the gradual change of the crystal structure from the {\it Pbnm}
symmetry for low Sr concentrations to the {\it Ibmm} symmetry for $x>0.5$.
(The (3,3,3), (3,3,1), and (2,2,5) peaks are allowed in the {\it Pbnm}
space group.)
These two structures exhibit different patterns of the tilts of MnO$_6$
octahedra \cite{NdSr,Vogt}. The {\it Pbnm} symmetry is normally found in the 
FM and CE-type states in the manganites \cite{Review}, while the {\it
Ibmm} symmetry is characteristic to the layered A-type AFM state in
Nd$_{1-x}$Sr$_x$MnO$_3$. The structural features characteristic to the latter
lattice symmetry therefore favor the A-type state with the associated
orbital and magnetic order. The gradual conversion of the {\it Pbnm} 
structure to the {\it Ibmm} structure with increasing $x$ can proceed in
a variety of possible ways. Spatial phase separation can, for example, 
take place. A more interesting scenario of a uniform average structure
exhibiting fluctuating correlated {\it Pbnm} domains is also plausible.
We note that our study, as well as the results of Refs. \cite{NdSr,Vogt}, do
not support the presence of phase separation for T$>$200 K.  
Independent on the exact microscopic structure of our sample, 
the pronounced anisotropy of the 
{\it Pbnm}-type correlations shows that in the correlated
domains, the coupling in the
$ab$ planes is much stronger than the interplane coupling. 
Since the $ab$ planes form the ferromagnetic layers in the A-type AFM state
for $x>$0.5, it is likely that the diminished interlayer coupling found in
the $x$=0.5 compound reflect the incipient A-type layered orbital order.
In the above scenario, therefore, the presence of the anisotropic lattice
fluctuations provide an {\it indirect} evidence of the local
layered orbital order
in the PM and the FM phases. 

Our preliminary diffraction
measurements on Nd$_{0.7}$Sr$_{0.3}$MnO$_3$ samples indicate the presence of
weak Bragg peaks characteristic to the {\it Pbnm} symmetry group.
To determine which of the two scenarios described above is correct, however,
it is
necessary to perform detailed structural studies of the 
Nd$_{1-x}$Sr$_x$MnO$_3$ samples for a number of Sr concentration in the
vicinity of $x$=0.5. In particular, the possible 
structural fluctuations should be
investigated in the AFM state itself and in its immediate vicinity.
This work is currently in progress.
 
We believe that
the newly observed type of structural correlations
cannot be attributed to small
admixtures of the A-type AFM or of the charge-ordered phases with the PI or
the FM phases for three reasons. Firstly,
at temperatures higher
than T=200 K, Nd$_{0.5}$Sr$_{0.5}$MnO$_3$ samples are believed to consist of
a single structural phase \cite{NdSr,Vogt}. Secondly,
the data of Fig. \ref{fig3} do not show any obvious anomaly
at T$_{AF}$. Finally, the (0, 0, 1)-type scattering abruptly disappears
below T$_{CO}$. 
We conclude that these
correlations are therefore intrinsic features of the PI
and the FM phases. 

Finally, we briefly discuss the x-ray diffuse scattering 
observed in the vicinity of the
main Bragg peaks. Such scattering is clearly visible in the data of 
Fig. \ref{fig2} as an elongated oval shape around the Bragg peak in the top
panel, and as a ``butterfly-like'' shape in the bottom panel. It has been
previously demonstrated that diffuse scattering of this kind
results from the presence of
uncorrelated lattice polarons \cite{Polarons}. The symmetry of the diffuse 
scattering in Fig. \ref{fig2} is consistent with the measurements reported
in Refs. \cite{Polarons,Corr},
and we therefore also interpret its intensity as reflecting the number of single
polarons present in the sample. 

\begin{figure}
\centerline{\epsfxsize=2.9in\epsfbox{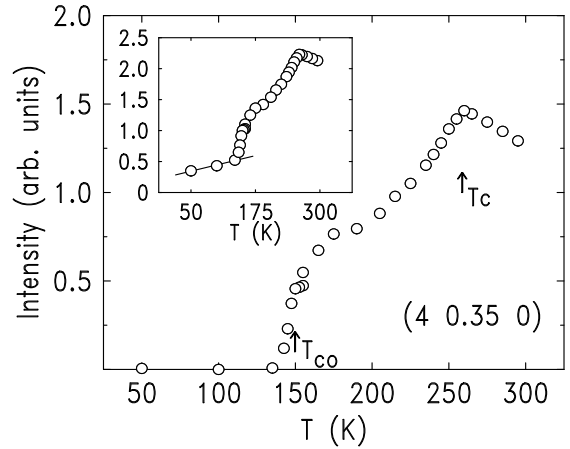}}
\vskip 5mm
\caption{Temperature dependence of the single-polaron diffuse scattering at
a scattering vector $Q$=(4, 0.35, 0). 
The estimated phonon contribution,
which is shown as a solid line in the inset,
is subtracted. The inset shows the raw data.}
\label{fig4}
\end{figure}

Fig. \ref{fig4} shows the temperature dependence
of this single-polaron diffuse scattering at a scattering vector 
$Q$=(4, 0.35, 0), that is on the
wing of the oval shape in the top panel of Fig. \ref{fig2}. The phonon 
contribution to the diffuse scattering was estimated from the data for
T$<$T$_{CO}$ \cite{Warren} and
subtracted. The remaining part, which reflects the polaron contribution,
shows a number of interesting features.
First, the number of polarons grows with decreasing temperature in the PI
phase, in agreement with Refs. \cite{Polarons,Corr}. 
At T$_c$, the intensity turns around and starts to fall.
At T$_{CO}$, the uncorrelated polarons abruptly disappear--
as would be expected following the formation of long-range orbital 
order. We note that the concentration of single
polarons does {\it not} exhibit an abrupt drop  at T$_c$, and the polarons are
clearly present in the FM phase.           
In contrast to this behavior, polaron scattering intensity abruptly decreases
upon the transition to the FM state in 
La$_{0.7}$Ca$_{0.3}$MnO$_3$ and
La$_{1.8}$Sr$_{1.8}$Mn$_2$O$_7$, compounds which are believed to
possess a microscopically uniform FM phase \cite{Polarons,Corr}.  
In Nd$_{0.5}$Sr$_{0.5}$MnO$_3$, {\it both} the 
uncorrelated polarons and the correlated anisotropic lattice distortions
are present below the Curie
temperature, and therefore the FM phase of this material is clearly less
uniform than the FM phase of the former compounds (see also Ref. \cite{AF}).
Unlike the intensity of the (3, 3, 3)
peak, the intensity of the single-polaron scattering shows a clear anomaly
at T$_c$. 
It appears, therefore, that the anisotropic lattice distortions discussed
above are not directly coupled with the system of the disordered lattice
polarons.

In summary, we report observation of anisotropic short-range correlated
structural distortions 
in the paramagnetic insulating phase and 
the ferromagnetic metallic phase of
Nd$_{0.5}$Sr$_{0.5}$MnO$_3$. 
We argue that the observed structural correlations indicate the presence 
of local layered orbital order in our samples.
Together with the anisotropy of the
magnetic excitations reported in Ref. \cite{AF}, these structural 
inhomogeneities show that the ferromagnetic and paramagnetic phases of
Nd$_{0.5}$Sr$_{0.5}$MnO$_3$ are nonuniform and exhibit large structural
and magnetic fluctuations which, possibly, reflect the incipient layered
magnetic order characteristic to the Nd$_{1-x}$Sr$_x$MnO$_3$ samples
with $x>0.51$.
Combined with the
results of previous work, our observations strongly
suggest that short-range structural correlations 
associated with local regions of orbital and magnetic order
play an important role 
in magnetoresistive manganite materials.

We are grateful to M. Croft, D. Gibbs, C. S. Nelson, and M. v. Zimmermann
for important discussions. This work was supported by the NSF under Grant
DMR-9802513, by Rutgers University, and by the US Department of
Energy under contract AC02-98CH10886.

%%%%%%%%%%%%%%%%%%%%%%%%%%%%%%%%%%%%%%%%%%%%%%%%%%%%%%%%%%%%%%%%%%%%%%%%

%%==============================================================================
%%==============================================================================

%%==============================================================================
%%==============================================================================

%%==============================================================================
%%==============================================================================

%%==============================================================================
%%==============================================================================


\begin{references}
\bibitem{Abs} One possible reason for the absence of the (4.5, 2.5, 0) and
(3.5, 1.5, 0) reflections in the data of Fig. 2 
could be that the displacements of atoms in the correlated domains
are predominantly transverse
(perpendicular to the (001) direction):
for a small displacement, $\bf\delta$
of an atom from its average position, 
the x-ray intensity is proportional to $|{\bf Q\cdot\delta}|^2$,
where {\bf Q} is the scattering vector.
\end{references}
\end{document}